\documentclass[prl,twocolumn,fleqn]{revtex4-1}
 \def\letter{1}\def\pr{1}
\usepackage{epsfig}
\usepackage{epstopdf}
\usepackage{graphicx}
\usepackage{ifthen}

\usepackage{amsmath}
\usepackage[psamsfonts]{amssymb}
\usepackage{euscript}

\usepackage{latexsym}
\usepackage[arrow,matrix,curve]{xy}

\newskip\humongous \humongous=0pt plus 1000pt minus 1000pt

\newif\ifdtup

\def\,{\hspace{-.1cm}}
\def\hsp{,\hspace{.7cm}}

\def\fc#1#2 {\frac{n}{q}#1\frac{n}{q}#2}

\newcommand{\vac}{\ensuremath{|0\rangle}}

\renewcommand{\theequation}{\arabic{section}.\arabic{equation}}
\renewcommand{\(}{\begin{equation}}
\renewcommand{\)}{end{equation} \vspace{-.05in}\linebreak}

\newcounter{saveeqn}
\newcounter{savealpheqn}

\newcommand{\alpheqn}{\setcounter{saveeqn}{\value{equation}}%
  \stepcounter{saveeqn}\setcounter{equation}{0}%
  \renewcommand{\theequation}{\mbox{\arabic{section}.\arabic{saveeqn}
\alph{equation}}}
  \renewcommand{\)}{\end{equation}}}
\def\part#1{\frac{\partial}{\partial{#1}}}%
\def\group#1{\refstepcounter{equation}\setcounter{saveeqn}
 {\value{equation}}%
  \label{#1}\setcounter{equation}{0}%
\renewcommand{\theequation}{\mbox{\arabic{section}.\arabic{saveeqn}
\alph{equation}}}
  \renewcommand{\)}{\end{equation}}}
\newcommand{\reseteqn}{\setcounter{equation}{\value{saveeqn}}%
  \renewcommand{\theequation}{\arabic{section}.\arabic{equation}}%
  \renewcommand{\)}{\end{equation}}}

\newcommand{\aalpheqn}{\setcounter{saveeqn}{\value{equation}}%
  \stepcounter{saveeqn}\setcounter{equation}{0}%
  \renewcommand{\theequation}{\mbox{
        \Alph{subsection}.\arabic{saveeqn}\alph{equation}}}
   \renewcommand{\)}{\end{equation}}}
\newcommand{\areseteqn}{\setcounter{equation}{\value{saveeqn}}%
  \renewcommand{\theequation}{\Alph{subsection}.\arabic{equation}}%
  \renewcommand{\)}{\end{equation}}}

\renewcommand{\thefootnote}{\alph{footnote}}
\renewcommand{\(}{\begin{equation}}
\renewcommand{\)}{\end{equation}}
\newcommand{\ba}{\begin{eqnarray}}
\newcommand{\ea}{\end{eqnarray}}
\newcommand{\cbp}{\mathop{\vtop{\ialign{##\crcr
   $\hfil\displaystyle{}\hfil$\crcr\noalign{\kern-13pt\nointerlineskip}
   \BIG{)}\hskip0pt\crcr\noalign{\kern3pt}}}}}
\newcommand{\pa}{\mathop{\vtop{\ialign{##\crcr

$\hfil\displaystyle{\oplus}\hfil$\crcr\noalign{\kern+1pt\nointerlineskip
}
   \hspace{.08in}$^{\alpha=0}$\hskip6pt\crcr\noalign{\kern3pt}}}}}
\renewcommand{\hsp}{,\hspace{.3in}}
\newcommand{\p}{^\prime}
\newcommand{\pp}{^{\prime\prime}}

\catcode`\@=11
\def\vereq#1#2{\lower3pt\vbox{\baselineskip1.5pt \lineskip1.5pt
\ialign{$\m@th#1\hfill##\hfil$\crcr#2\crcr\sim\crcr}}}
\catcode`\@=12

\renewcommand{\(}{\begin{equation}}
\renewcommand{\)}{\end{equation}}

\def\pin#1{\int \frac{d#1}{2\pi}}
\def\pink#1{\int \frac{d^{#1}k}{(2\pi)^{#1}}}

\def\pinkp#1{\int \frac{d^{#1}k\p}{(2\pi)^{#1}}}

\def\Bd#1{B^\dag_{k_{#1}}}

\def\df{\mathcal{D}_f}

\def\I{\mathcal{I}}

\newcommand{\beas}{\begin{eqnarray*}}
\newcommand{\eeas}{\end{eqnarray*}}

\newcommand{\bquo}{\begin{quote}}
\newcommand{\enqu}{\end{quote}}

\def\ch{{\mathcal{H}}}

\def\okp#1{\omega_{k\p_{#1}}}
\def\V#1{V^{(#1)}[gf(x)]}

\newcommand{\beq}{\begin{equation}}
\newcommand{\eeq}{\end{equation}}
\newcommand{\bea}{\begin{eqnarray}}
\newcommand{\eea}{\end{eqnarray}}

\newskip\humongous \humongous=0pt plus 1000pt minus 1000pt

\newif\ifdtup

\catcode`\@=11

\ifthenelse{\equal{\letter}{0}}{ 

\def\theequation{\arabic{section}.\arabic{equation}}

\def\@normalsize{\@setsize\normalsize{15pt}\xiipt\@xiipt
\abovedisplayskip 14pt plus3pt minus3pt%
\belowdisplayskip \abovedisplayskip
\abovedisplayshortskip \z@ plus3pt%
\belowdisplayshortskip 7pt plus3.5pt minus0pt}

\def\small{\@setsize\small{13.6pt}\xipt\@xipt
\abovedisplayskip 13pt plus3pt minus3pt%
\belowdisplayskip \abovedisplayskip
\abovedisplayshortskip \z@ plus3pt%
\belowdisplayshortskip 7pt plus3.5pt minus0pt
\def\@listi{\parsep 4.5pt plus 2pt minus 1pt
      \itemsep \parsep
      \topsep 9pt plus 3pt minus 3pt}}

\relax

\def\section{\@startsection{section}{1}{\z@}{3.5ex plus 1ex minus  .2ex}{2.3ex plus .2ex}{\large\bf}}

\def\thesection{\arabic{section}}
\def\thesubsection{\arabic{section}.\arabic{subsection}}

\def\appendix{\setcounter{section}{0}
 \def\thesection{Appendix \Alph{section}}
 \def\thesubsection{\Alph{section}.\arabic{subsection}}
 \def\theequation{\Alph{section}.\arabic{equation}}}
\renewcommand{\theequation}{\arabic{section}.\arabic{equation}}

}{
\renewcommand{\theequation}{\arabic{equation}}

} 

\begin{document}
\def\thefootnote{\fnsymbol{footnote}}
\def\thetitle{An Alternative to Collective Coordinates}
\def\autone{Jarah Evslin}
\def\auttwo{Hengyuan Guo}
\def\affa{Institute of Modern Physics, NanChangLu 509, Lanzhou 730000, China}
\def\affb{University of the Chinese Academy of Sciences, YuQuanLu 19A, Beijing 100049, China}

\title{Titolo}

\ifthenelse{\equal{\pr}{1}}{
\title{\thetitle}
\author{\autone}
\author{\auttwo}
\affiliation {\affa}
\affiliation {\affb}

}{

\begin{center}
{\large {\bf \thetitle}}

\bigskip

\bigskip


{\large \noindent  \autone{${}^{1,2}$} and \auttwo{${}^{1,2}$}}


\vskip.7cm

1) \affa\\
2) \affb\\

\end{center}

}

\begin{abstract}
\noindent

\noindent
Collective coordinates provide a powerful tool for separating collective and elementary excitations, allowing both to be treated in the full quantum theory.  The price is a canonical transformation which leads to a complicated starting point for subsequent calculations.  Sometimes the collective behavior of a soliton is simple but nontrivial, and one is interested in the elementary excitations.  We show that in this case an alternative prescription suffices, in which the canonical transformation is not necessary.  The use of a nonperturbative operator which creates a soliton state allows the theory to be constructed perturbatively in terms of the soliton normal modes.  We show how translation invariance may be perturbatively imposed.  We apply this to construct the two-loop ground state of an arbitrary scalar kink.

\end{abstract}

%
\setcounter{footnote}{0}
\renewcommand{\thefootnote}{\arabic{footnote}}

\ifthenelse{\equal{\pr}{1}}{
\maketitle
}{}

When a theory is reformulated in terms of collective coordinates, some phenomena involving large numbers of elementary quanta, such as plasma waves, can be treated in perturbation theory \cite{bp1}.  Two groups applied collective coordinates to quantum solitons in the fateful Summer of 1975, allowing a treatment of the scattering of quantum solitons.  In \cite{gjs}, following the spirit of \cite{bp1}, the collective coordinates are related to the elementary fields by a canonical transformation.  This transformation allows a straightforward quantization of the system.  However it comes with a price, the theory becomes rather complicated and power counting renormalizability is lost.  Nevertheless, the authors are still able to study a soliton in motion and in Refs. \cite{vega,verwaest} the two-loop correction to a soliton energy is reproduced.  A functional integral formulation is employed to avoid the complications of quantum states.  

In \cite{christlee75} collective coordinates are introduced without the canonical transformation.   Following \cite{bp1}, the formulation was Hamiltonian.  As no transformation was used, the authors are forced to quantize the theory without collective coordinates to determine the operator ordering in the theory with collective coordinates.   The theory is still ``considerably more complex than that usually encountered in quantum field theory'' but is now simple enough that the authors can treat two soliton scattering.  Due to the complexity of both approaches, quantum states were never considered beyond one loop, where the theories are sums of uncoupled quantum harmonic oscillators.

Sometimes one is not interested in the collective excitations.  For example, one may be interested in the quantum structure of a soliton in its rest frame.  After all, intuition from large $N$ \cite{wittenbar} suggests that hadrons are quantum solitons and so their masses, form factors and general matrix elements may be calculated by solving for the corresponding quantum state.  In this case, we will propose a much simpler alternative to collective coordinates which allows one to pass to higher numbers of loops using reasonably elementary computations.

For concreteness we will describe our formalism \cite{memassa,me2stato} in the case of a real scalar field theory in 1+1 dimensions, described by the Hamiltonian
\bea
H&=&\int dx \ch(x) \label{hd}\\
\ch(x)&=&\frac{1}{2}:\pi(x)\pi(x):_a+\frac{1}{2}:\partial_x\phi(x)\partial_x\phi(x):_a\nonumber\\
&&+\frac{1}{g^2}:V[g\phi(x)]:_a\nonumber
\eea
where $::_a$ is the normal ordering defined below.  Let
\beq
\phi(x,t)=f(x) \label{fd}
\eeq
be a kink solution to the classical equations of motion.  We will always work in the Schrodinger picture.  

We assume that $V\pp[gf(-\infty)]=V\pp[gf(\infty)]$ and define $M^2/2$ to be equal to this value.  Here the prime denotes a functional derivative of $V$ with respect to it argument.

As (\ref{fd}) is a solution of the classical equations of motion, we might be tempted to expand the quantum field as $\phi(x)=f(x)+\eta(x)$.  Then $\phi\rightarrow\eta=\phi-f$ would be a passive transformation of the fields.  After this transformation, the quadratic part of the Hamiltonian $H[\eta]$ would describe small perturbations about the classical kink, and one could proceed perturbatively.

Instead of this passive transformation of the fields, following the standard approach \cite{dhn2,rajaraman}, we will consider an active transformation of the functionals acting on the fields.  In particular, we transform the Hamiltonian
\beq
H[\phi,\pi]\rightarrow H\p[\phi,\pi]=H[f+\phi,\pi].
\eeq
Below we will perform the same transformation on the momentum operator $P$.  The new observation that lies behind our approach is that $H\p$ and $H$ are unitarily equivalent, because
\beq
H\p=\df^\dag H\df \label{hpd}
\eeq
where we have defined the translation operator
\beq
\df={\rm{exp}}\left(-i\int dx f(x)\pi(x)\right). \label{df}
\eeq
In general Eq.~(\ref{hpd}) will be applied to the regularized and renormalized $H$ and will be our definition of the regularized and renormalized $H\p$.  This eliminates the need to separately regularize $H\p$ and then to guess the correct regulator matching condition to apply when both regulators are taken to infinity.  It has long been known \cite{rebhan} that the dependence on the unknown matching condition leads to wrong answers in otherwise correct calculations.  In (\ref{hd}) all UV divergences are removed by the normal ordering, but this choice was not necessary for our approach.

The unitary equivalence (\ref{hpd}) implies that $H$ and $H\p$ have the same spectrum.  Therefore the vacuum and the kink ground state are eigenstates of both Hamiltonians, with the same eigenvalues.  We are then free to use the vacuum Hamiltonian $H$ to calculate the vacuum energy and the kink Hamiltonian $H\p$ to calculate the kink ground state energy.  We will argue that this choice allows both calculations to be performed in perturbation theory.

This procedure will give us not only the energies of the kink states, but also the kink states themselves.  Once an eigenstate of $H\p$ is found, one need only apply $\df$ to arrive at the corresponding $H$ eigenstate.  For example, if $\vac$ is the eigenstate of $H\p$ corresponding to the kink ground state, then $\df\vac$ is the corresponding eigenstate $|K\rangle$ of $H$.  

This correspondence works already at tree level.  Let $|\Omega\rangle$ be a free vacuum of $H$ that satisfies
\beq
\langle \Omega|\phi(x)|\Omega\rangle=0. \label{ff0}
\eeq
This can be arranged by shifting $\phi$ by a constant.  Then $\df^\dag|\Omega\rangle$ is the free vacuum as an eigenstate of $H\p$. 

On the other hand, $|\Omega\rangle$ is not eigenstate of $H\p$, or even of its free part.  However it has a vanishing form factor  (\ref{ff0}) which one may expect for a tree-level vacuum.  The corresponding state $\df|\Omega\rangle$ in the eigenbasis of $H$ is obtained via the unitary transformation.  As a result of (\ref{ff0}) it has a form factor which reproduces the classical kink profile
\beq
\langle \Omega|\df^\dag\phi(x)\df|\Omega\rangle=f(x). \label{fff}
\eeq
The state $\df|\Omega\rangle$ is not the kink ground state $|K\rangle$, indeed it is not even an eigenstate of $H$ just as $|\Omega\rangle$ is not an eigenstate of $H\p$.   However it has the correct form factor (\ref{fff}),  leading one to suspect that the difference between the two can be calculated in perturbation theory as we now describe.   

We have argued that the eigenstate $\vac$ of $H\p$ corresponding to the kink ground state is close to $|\Omega\rangle$.  Our goal in this note will be to obtain a procedure which provides successively better approximations to $\vac$.

The corresponding eigenstate of $H$ will be
\beq
|K\rangle=\df \vac. \label{kdef}
\eeq
The eigenvalue equation
\beq
H\p\vac=Q\vac
\eeq
is easily solved at leading order as it reduces to a free theory and subleading orders can be solved by simply fixing higher order coefficients, and so it is in principle possible to find an all-orders solution for $\vac$.  To obtain the correct eigenstate, we fix the leading order energy to be minimal among eigenstates of the free part of $H\p$.  Had we not performed the unitary transformation, this program would have failed already at the leading order, due to the inverse coupling appearing in the leading term in the soliton mass.

To perform this perturbative calculation, we first expand $H\p$ in powers of the coupling
\bea
H\p&=&\df^\dag H\df=Q_0+\sum_{n=2}^{\infty}H_n\\
H_2&=&\frac{1}{2}\int dx\left[:\pi^2(x):_a+:\left(\partial_x\phi(x)\right)^2:_a\right.\nonumber\\
&&\left.+V^{\prime\prime}[gf(x)]:\phi^2(x):_a\right.].\nonumber
\eea
$Q_0$ is the classical kink mass and $H_n$ is order $g^{n-2}$.  

At one loop, only $H_2$ is relevant.  The constant frequency $\omega$ solutions of its classical equations of motion are continuum normal modes $g_k(x)$ with $\omega_k=\sqrt{M^2+k^2}$, discrete breathers and a Goldstone mode $g_B(x)= f^\prime(x)/\sqrt{Q_0}$.  Note that the definition of $\omega_k$ fixes the parametrization of $k$ up to a sign.  For brevity of notation, we will not distinguish between continuum solutions and breathers, and so it will be implicit that integrals over the continuous variable $k$ include a sum over the breathers, and $2\pi$ times a Dirac delta function of continuum $k$ should be understood as a Kronecker delta of breathers.
 
We choose the normalization conditions
\beq
\int dx g_{k_1} (x) g^*_{k_2}(x)=2\pi \delta(k_1-k_2),\ 
\int dx |g_{B}(x)|^2=1
\eeq
and conventions
\beq
g_k(-x)=g_k^*(x)=g_{-k}(x),\ \tilde{g}(p)=\int dx g(x) e^{ipx}
\eeq
leading to the completeness relations
\beq
g_B(x)g_B(y)+\pin{k}g_k(x)g^*_{k}(y)=\delta(x-y). \label{comp}
\eeq

As it is independent of time, the Schrodinger picture field $\phi(x)$ may be expanded in any basis of functions even in the full, interacting theory.  We will expand it in terms of plane waves
\bea
\phi(x)&=&\pin{p}\left(A^\dag_p+\frac{A_{-p}}{2\omega_p}\right) e^{-ipx}\\
 \pi(x)&=&i\pin{p}\left(\omega_pA^\dag_p-\frac{A_{-p}}{2}\right) e^{-ipx}
\nonumber
\eea
and also normal modes \cite{cahill76}
\bea
\phi(x)&=&\phi_0 g_B(x) +\pin{k}\left(B_k^\dag+\frac{B_{-k}}{2\omega_k}\right) g_k(x)\\
\pi(x)&=&\pi_0 g_B(x)+i\pin{k}\left(\omega_kB_k^\dag - \frac{B_{-k}}{2}\right) g_k(x).\nonumber
\eea
Define the plane wave (normal mode) normal ordering $::_a$ ($::_b$) by moving all $A^\dag$ (all $\phi_0$ and $B^\dag$) to the left.  The canonical algebra obeyed by $\phi(x)$ and $\pi(x)$ then implies
\bea
[A_p,A_q^\dag]&=&2\pi\delta(p-q)\\
{[\phi_0,\pi_0]}&=&i\hsp
[B_{k_1},B^\dag_{k_2}]=2\pi\delta(k_1-k_2).\nonumber
\eea

Decomposing fields in terms of the plane wave operators, Bogoliubov transforming to the normal mode operators and then normal mode normal ordering one finds that the one-loop Hamiltonian is a sum of quantum harmonic oscillators plus a free quantum mechanical particle for the center of mass
\bea
H_2&=&Q_1+\frac{\pi_0^2}{2}+\pin{k}\omega_k B^\dag_k B_k \\
Q_1&=&-\frac{1}{4}\pin{k}\pin{p}\frac{(\omega_p-\omega_k)^2}{\omega_p}\tilde{g}^2_{k}(p)\nonumber\\
&&-\frac{1}{4}\pin{p}\omega_p\tilde{g}_{B}(p)\tilde{g}_{B}(p)\nonumber
\eea
where $Q_1$ is the one-loop kink mass.  The one-loop kink ground state $\vac_0$ is therefore the solution of
\beq
\pi_0\vac_0=B_k\vac_0=0. \label{v0}
\eeq
The whole spectrum may be obtained exactly at one-loop by creating normal modes with $B^\dag_k$ and boosting with $e^{i\phi_0 k}$.  The state $|0\rangle_0$ is the first term in the semiclassical expansion in powers of $\sqrt{\hbar}$
\beq
\vac=\sum_{i=0}^\infty |0\rangle_{i} \label{semi}
\eeq
where the $n$-loop ground state is the sum up to $i=2n-2$. 

We will now consider the construction of the ground state $|K\rangle$ at higher orders.  As the one-loop spectrum is known exactly, the generalization of what follows to other states is trivial.  Recall that, using (\ref{kdef}), it is sufficient to construct $\vac$.  $|K\rangle$ is annihilated by the momentum operator
\beq
P=-\int dx \pi(x)\partial_x \phi(x). \label{pdef}
\eeq
Therefore $\vac$ is annihilated by its unitary transform
\beq
P\p=\df^\dag P\df=P-\sqrt{Q_0}\pi_0.
\eeq
As $g$ has dimensions of [action]${}^{-1/2}$, the quantity $g\hbar^{1/2}$ is dimensionless.  Setting $\hbar$ to unity, the semiclassical expansion in $\hbar$ is therefore equivalent to an expansion in $g$.  While $P$ and $\pi_0$ are independent of $g$, $\sqrt{Q_0}$ is proportional to $g^{-2}$ and so $\hbar^{-1}$.  Thus the action of $P$ preserves the order in the semiclassical expansion while $\sqrt{Q_0}\pi_0$ reduces the order by one.  Therefore 
\beq
\left(P-\sqrt{Q_0}\pi_0\right)\vac=0\label{pcon}
 \eeq
implies the recursion relation
\beq
P|0\rangle_i=\sqrt{Q_0}\pi_0|0\rangle_{i+1}. \label{ti}
\eeq
Up to the kernel of $\pi_0$, this determines order $i+1$ states from order $i$ states.  

We can now state the critical difference between our approach and the collective coordinate approach.  Whereas the collective coordinate approach imposes translation invariance exactly, we only solve the recursion relation (\ref{ti}) up to the order at which we intend to find the state.  As a result, no nonlinear canonical transformation is required, only the linear Bogoliubov transformation that relates the $A_p$ and $B_k$.  Thus we do not arrive at a complicated Hamiltonian.  {\it{On the contrary, perturbation theory is greatly simplified as we only need to solve for components in the kernel of $\pi_0$, the rest of the state is fixed by the recursion relation.}}

The momentum operator (\ref{pdef}) is
\bea
P&=&\pin{k}\Delta_{kB}\left[i\phi_0 \left(-\omega_kB_k^\dag+\frac{B_{-k}}{2}\right)\right.\\
&&\left.+\pi_0\left(B_k^\dag+\frac{B_{-k}}{2\omega_k}\right)\right]\nonumber\\
&&+i\pink{2}\Delta_{k_1k_2}\left(-\omega_{k_1}B_{k_1}^\dag B_{k_2}^\dag\right.\nonumber\\
&&\left.+\frac{B_{-k_1}B_{-k_2}}{4\omega_{k_2}}-\frac{1}{2}\left(1+\frac{\omega_{k_1}}{\omega_{k_2}}\right)B^\dag_{k_1}B_{-k_2}
\right)\nonumber
\eea
where we have defined the  matrix
\beq
\Delta_{ij}=\int dx g_i(x) g\p_j(x).
\eeq
Integration by parts, using the fact that all $g_i(x)$ vanish asymptotically, exchanges the indices and introduces a minus sign, so $\Delta_{ij}$ is antisymmetric.  We can expand the $i$th order kink ground state as
\bea
\vac_i&=& Q_0^{-i/2}\sum_{m,n=0}^\infty\pink{n}\gamma_i^{mn}(k_1\cdots k_n)\nonumber\\
&&\times \phi_0^m\Bd1\cdots\Bd n\vac_0. \label{gameq}
\eea
Then the recursion relation becomes
\bea
&&\gamma_{i+1}^{mn}(k_1\cdots k_n)=\Delta_{k_n B}\left(\gamma_i^{m,n-1}(k_1\cdots k_{n-1})\right.\nonumber\\
&&\left.+\frac{\omega_{k_n}}{m}\gamma_i^{m-2,n-1}(k_1\cdots k_{n-1})\right)
 \nonumber\\
&&+(n+1)\pin{k\p}\Delta_{-k\p B}\left(\frac{\gamma_i^{m,n+1}(k_1\cdots k_n,k\p)}{2\omega_{k\p}}\right.\nonumber\\
&&\left.
-\frac{\gamma_i^{m-2,n+1}(k_1\cdots k_n,k\p)}{2m}\right)\nonumber\\
&&+\frac{\omega_{k_{n-1}}\Delta_{k_{n-1}k_n}}{m}\gamma_i^{m-1,n-2}(k_1\cdots k_{n-2})\nonumber\\
&&+\frac{n}{2m}\pin{k\p}\Delta_{k_n,-k\p}\,\left(\,1+\frac{\omega_{k_n}}{\omega_{k\p}}\,\right)\,\gamma^{m-1,n}_i(k_1\cdots k_{n-1},k\p)
\nonumber\\
&&-\frac{(n+2)(n+1)}{2m}\int\frac{d^2k\p}{(2\pi)^2}\frac{\Delta_{-k\p_1,-k\p_2}}{2\omega_{k\p_2}}\nonumber\\
&&\times \gamma_i^{m-1,n+2}(k_1\cdots k_{n},k\p_1,k\p_2).
\label{rrs}
\eea
We have assumed here that $\gamma_i$ is symmetric under a permutation of the $k_j$, but (\ref{rrs}) yields a $\gamma_{i+1}$ which is not symmetric.  Therefore, before each successive application of the recursion relation, it is necessary to symmetrize $\gamma_{i+1}$.  The definition of the state (\ref{gameq}) is invariant under this symmetrization.

As is, the recursion relation applies to any kink state whose center of mass is at rest.  To restrict to the ground state, we need only impose the initial condition
\beq
\gamma_0^{mn}=\delta_{m0}\delta_{n0}\gamma_0^{00}.
\eeq
One recursion yields
\bea
\gamma_1^{12}(k_1,k_2)=\frac{\left(\omega_{k_1}-\omega_{k_2}\right)\Delta_{k_1k_2}}{2}\gamma_0^{00}\nonumber\\
\gamma_1^{21}(k_1)=\frac{\omega_{k_1}\Delta_{k_1B}}{2}\gamma_0^{00}. \label{g121}
\eea
Two yield the two-loop state up to the kernel of $\pi_0$, corresponding to $\gamma_2^{0n}$.  These are reported in Ref.~\cite{colcor}.

The terms $\gamma_2^{0n}$, which are in the kernel of $\pi_0$ can be found using ordinary perturbation theory as follows.

First define $\Gamma$ to be any solution of
\bea
&&\sum_{j=0}^i \left(H_{i+2-j}-Q_{\frac{i-j}{2}+1}\right)\vac_j\label{scheq}\\
&&=\sum_{mn} \pink{n} \Gamma_i^{mn}(k_1\cdots k_n)\phi_0^m B_{k_1}^\dag\cdots B_{k_n}^\dag\vac_0\nonumber
\eea
where $\Gamma_i$ is of order $O(g^i)$.   Recall that $\vac_j$ is determined by $\gamma_j$ and so $\Gamma$ is a function of $\gamma$.  Then observe that the Schrodinger Equation
\beq
(H-Q)\vac=0
\eeq
is solved by any $\gamma$ such that
\beq
\Gamma_i^{mn}=0. \label{g0}
\eeq
Recall that only the $\gamma_i^{0n}$ need be determined perturbatively, as only they lie in the kernel of $\pi_0$.  The other components were already fixed by the recursion relation (\ref{ti}).  

To solve (\ref{scheq}) we first note that
\beq
H_n=\frac{1}{n!}\int dx \V n:\phi^n(x):_a
\eeq
where $\V{n}$ is the $n$th derivative of $g^{n-2}V[g\phi(x)]$ evaluated at $\phi(x)=f(x)$.   These are converted into normal mode normal ordered expressions using the Wick's theorem stated and proved in \cite{wick}.  As normal mode normal ordered expressions act simply on $\vac_0$, one can easily use Eq.~(\ref{scheq}) to write $\Gamma$ in terms of $\gamma$.  At each new order $i$, the $\gamma_i$ appear linearly and so the condition that $\Gamma_i=0$ in (\ref{g0}) is uniquely solved for $\gamma_i$.

The usual IR problems associated to perturbation theory in the presence of a continuous spectrum are resolved here by the momentum constraint (\ref{pcon}), as they are resolved in the case of the collective coordinate approach.  As this perturbative calculation is standard, it is reported in the companion paper \cite{colcor}.  It yields a general formula valid for the energy of any scalar kink at two loops
\bea
Q_2
&=&\frac{V_{\I\I}}{8}-\frac{1}{8}\pin{k\p}\frac{\left|V_{\I k\p}\right|^2}{\okp{}^2}\nonumber\\
&&-\frac{1}{48}\pinkp{3} \frac{\left|V_{k\p_1k\p_2k\p_3}\right|^2}{\omega_{k\p_1}\omega_{k\p_2}\omega_{k\p_3}\left(\omega_{k\p_1}+\omega_{k\p_2}+\omega_{k\p_3}\right)}\nonumber\\
&&  +\frac{1}{16Q_0}\pinkp{2}\frac{\left|\left(\omega_{k_1\p}-\omega_{k_2\p}\right)\Delta_{k_1\p k_2\p}\right|^2}{\omega_{k\p_1}\omega_{k\p_2}}\nonumber\\
&&  -\frac{1}{8Q_0}\pin{k\p}
\left|f^{\prime\prime}(x)\right|^2 
\nonumber
\eea
where 
\beq
V_{\I\stackrel{m}{\cdots}\I,\alpha_1\cdots\alpha_n}=\int dx V^{(2m+n)}[gf(x)]\I^m(x) g_{\alpha_1}(x)\cdots g_{\alpha_n(x)}.
\eeq
Here we have introduced the contraction factor $\I(x)$ determined by \cite{wick}
\beq
\partial_x \I(x)=\pin{k}\frac{1}{2\omega_k}\partial_x\left|g_{k}(x)\right|^2 \label{di}
\eeq
and the condition that it vanish at infinity.

The two-loop scalar kink mass was previously only known in the Sine-Gordon case \cite{vega,verwaest}.  There it was derived from 13 UV divergent diagrams, which can be combined into five finite combinations.  Our terms are always each UV finite, as we have normal-ordered from the beginning.  In the Sine-Gordon case the terms in our energy formula are these five finite combinations.  Our formula on the other hand also applies to kinks in many other models, such as $\phi^{2n}$ models.

However, by finding the two-loop state, and not just the mass, one can do much more.  For example, it would be straightforward to calculate form factors \cite{kimform} and matrix elements.  This would allow, for the first time, a truly quantum approach to meson-kink scattering  \cite{adamscat,chris,wobble}, breather excitation, acceleration \cite{melac,melac2} and more.

\section* {Acknowledgement}

\noindent
JE is supported by the CAS Key Research Program of Frontier Sciences grant QYZDY-SSW-SLH006 and the NSFC MianShang grants 11875296 and 11675223.   JE also thanks the Recruitment Program of High-end Foreign Experts for support.

\end{document}